\documentclass[aps,prb,preprint,superscriptaddress]{revtex4-2}
\usepackage{dcolumn}   
\usepackage{bm}        
\usepackage{amsmath}
\usepackage{amssymb}
\usepackage[utf8]{inputenc}
\usepackage{graphicx,siunitx}
\usepackage{amsmath}
\DeclareUnicodeCharacter{2212}{-}
\usepackage{booktabs}
\usepackage{subcaption}
\usepackage[svgnames]{xcolor}
\usepackage[export]{adjustbox}
\usepackage[percent]{overpic}
\usepackage{array}
\usepackage[style=base]{caption}
\DeclareSIUnit\angstrom{\text {Å}}
\begin{document}
	\title{Pressure induced structural phase transition and magneto-elastic coupling in Eu doped LaCrO$_3$ }
	
	\author{Asish Kumar Mishra}
	\affiliation{Department of Physical Sciences, Indian Institute of Science Education and Research Kolkata, Mohanpur Campus, Mohanpur 741246, Nadia, West Bengal, India.}
	\affiliation{National Centre for High-Pressure Studies, Department of Physical Sciences, Indian Institute of Science Education and Research Kolkata, Mohanpur Campus, Mohanpur 741246, Nadia, West Bengal, India.}
	\author{Mrinmay Sahu}
	\affiliation{Department of Aerospace Engineering, Iowa State University, Ames, Iowa 50011, USA}
		
	\author{Bhagyashri Giri}
	\affiliation{Department of Physical Sciences, Indian Institute of Science Education and Research Kolkata, Mohanpur Campus, Mohanpur 741246, Nadia, West Bengal, India.}
	\affiliation{National Centre for High-Pressure Studies, Department of Physical Sciences, Indian Institute of Science Education and Research Kolkata, Mohanpur Campus, Mohanpur 741246, Nadia, West Bengal, India.}
		\author{Bidisha Mukherjee}
	\affiliation{IMPMC, Sorbonne Université, CNRS, MNHN, 4, place Jussieu, 75005 Paris, France}
		\author{Suvashree Mukherjee}
	\affiliation{Department of Physical Sciences, Indian Institute of Science Education and Research Kolkata, Mohanpur Campus, Mohanpur 741246, Nadia, West Bengal, India.}
	\affiliation{National Centre for High-Pressure Studies, Department of Physical Sciences, Indian Institute of Science Education and Research Kolkata, Mohanpur Campus, Mohanpur 741246, Nadia, West Bengal, India.}
		\author{Harekrishna Bhunia}
	\affiliation{Department of Physical Sciences, Indian Institute of Science Education and Research Kolkata, Mohanpur Campus, Mohanpur 741246, Nadia, West Bengal, India.}
	\author{Tamalkanti Mukherjee}
	\affiliation{Department of Geology and Geophysics, Indian Institute of Technology, Kharagpur-721302, West Bengal, India.}
		\author{Sujoy Ghosh}
	\affiliation{Department of Geology and Geophysics, Indian Institute of Technology, Kharagpur-721302, West Bengal, India.}
		\author{Partha Mitra}
	\affiliation{Department of Physical Sciences, Indian Institute of Science Education and Research Kolkata, Mohanpur Campus, Mohanpur 741246, Nadia, West Bengal, India.}
		\author{Peter Liermann}
	\affiliation{Photon Science, Deutsches Elektronen Synchrotron, 22607 Hamburg, Germany}
		\author{Goutam Dev Mukherjee}
 \email [Corresponding author:]{goutamdev@iiserkol.ac.in}
	\affiliation{Department of Physical Sciences, Indian Institute of Science Education and Research Kolkata, Mohanpur Campus, Mohanpur 741246, Nadia, West Bengal, India.}
	\affiliation{National Centre for High-Pressure Studies, Department of Physical Sciences, Indian Institute of Science Education and Research Kolkata, Mohanpur Campus, Mohanpur 741246, Nadia, West Bengal, India.}
		\date{\today}
		\begin{abstract}
			In this study, we have carried out a detailed high pressure investigation on 5$\%$ Eu doped LaCrO$_3$ (ELCO) using synchrotron X-ray diffraction (XRD), micro Raman spectroscopy, and low-temperature magnetization measurements to correlate the structural and magnetic properties under pressure. The high pressure XRD reveals the orthorhombic to rhombohedral structural phase transition around 10.6 GPa. The high pressure Raman data corroborate this result and indicate that the lattice instabilities associated with the low-frequency soft modes play a crucial role in driving this transition. In addition, a pronounced anomaly in the Raman shift and integrated intensity of several Raman modes is observed around 4.5 GPa. By combining the low-temperature magnetization measurements with the anomalies observed in the XRD and Raman data and by comparing with the similar results in LaCrO$_3$ (LCO), a pressure-induced change in the magnetic ground state to antiferromagnetic ordering is predicted.  
		\end{abstract}
		\maketitle
		\newpage
\section{introduction}
In recent years, the rare earth orthochromites with the general formula RCrO$_3$ (R= rare earth) have attracted significant research attention due to their wide variety of physical properties such as distorted crystal structure, complex magnetic behaviour, multiferroicity, spin reorientation transition, magneto structural coupling etc \cite{sahu2007rare,rajeswaran2012field,raveau2014impact,taheri2016magnetic,singh2018effect}. These tunable structural, electrical, and magnetic properties make these materials suitable for many technological applications such as solid oxide fuel cells, magneto optical devices, gas sensors, catalysts, etc \cite{sfeir2003lacro3,siemons2007preparation,kojima2002recent,lund2012composite,zwinkels1999preparation,arakawa1981catalytic}. These compounds crystallise in a distorted perovskite structure and the properties of these materials are sensitive to the crystal distortion, which increases with the decrease of rare earth ionic radius \cite{singh2018effect,late2016probing}. These materials exhibit canted antiferromagnetic ordering below their Neel temperature ($T_N$), and also show weak ferromagnetism arising from the antisymmetric Dzialoshinskii–Moriya (DM) interaction \cite{hornreich1978magnetic,rajeswaran2012field}. 

As a member of this family, LaCrO$_3$  crystallises in an orthorhombic structure and exhibits  G-type antiferromagnetic ordering below  $T_N$ $\sim$ 290 K \cite{tiwari2015magnetostructural,tiwari2013dielectric}. It is known to undergo a structural phase transition from orthorhombic to rhombohedral phase around 530 K \cite{hashimoto2000analysis,iliev2006raman,oikawa2000structural}. The orthorhombic to rhombohedral phase transition can also be obtained at room temperature under an applied pressure of approximately $5$ GPa \cite{hashimoto1998pressure,shibasaki2005exploration,bhadram2021pressure}. Zhou et al. reported that the G-type antiferromagnetic ordering does not collapse at the structural transition. The Neel temperature increases with the increase in pressure and exceeds 300 K above approximately 2.2 GPa pressure. In addition, a rotation of spin direction is observed at the structural phase transition \cite{zhou2011magnetic}. The physical properties of LaCrO$_3$ strongly depend on the structural distortion, Cr-O-Cr bond angle, t$_{2g}$-e$_g$ orbital hybridisation \cite{singh2018effect,zhou2010intrinsic}. Substitutions at the La or Cr site can effectively tune these parameters through ionic radius mismatch and modification of the rare earth and transition metal interaction, thereby affecting the structural, magnetic, and optical properties. Yimin et al. reported a reduction of the optical bandgap from 3.13 eV to 2.80 eV by increasing Ca concentration at the La site \cite{wan2019synthesis}. The G-type AFM ground state of LaCrO$_3$ changes to the C-type AFM state at 25$\%$ and 50$\%$ Ca doping, and at 75$\%$ Ca doping it becomes ferromagnetic \cite{swami2023onset}. The substitution of rare-earth ion Eu at the La site causes lattice distortion that reduces the hybridisation between O-2p and Cr-3d orbitals, which results in the reduction of the optical band gap. Furthermore, the anisotropic exchange interaction between the Cr$^{3+}$ and Eu$^{3+}$ moments leads to the reduction of total magnetic moment and magnetic saturation as reported by Siddique et al. \cite{siddique2021intrinsic}. Bouasla et al. investigated the optical and magnetic properties of LaCrO$_3$ by substituting rare earth ions and Ca at the La site and reported the highest optical conductivity for the Eu-doped samples. Their results revealed that the Neel temperature decreases with the decrease in the rare earth ionic radius, indicating a strong correlation between lattice distortion and the magnetic properties \cite{bouasla2023optical}. Pravin Kumar et al. reported the formation of Cr$^{6+}$ centres due to the substitution of isovalent Gd ion at the La site and found that the dc conductivity for the samples having Gd concentrations more than 10$\%$ is predominantly governed by the polarons created due to the Cr$^{6+}$ centres \cite{kumar2013effect}. All of these studies have focused mainly on the effect of chemical substitution on the physical properties of LaCrO$_3$ at ambient pressure conditions. However, the investigation on the effect of external pressure on the chemically substituted LaCrO$_3$ is surprisingly rare. The external pressure can effectively tune the bond angle, bond length and electronic interactions, thereby significantly modifying the structural, electronic and magnetic properties of the system without altering the chemical composition \cite{mukherjee2026probing,mukherjee2024pressure,mishra2026pressure,sahu2024high}. Therefore, investigating the combined effect of chemical pressure induced by the substitution of rare earth ion and external pressure is essential to understand the structure-property correlation in LaCrO$_3$. The rare earth ion Eu$^{3+}$ has a smaller ionic radius (1.28 \AA) as compared to La$^{3+}$ (1.36 \AA), so the substitution of Eu at the La site induces chemical pressure in the lattice and thereby modifies  the structural distortion. In LaCrO$_3$, the La$^{3+}$ is nonmagnetic, and the long-range magnetic ordering is primarily due to the Cr$^{3+}$-O- Cr$^{3+}$ super exchange interaction. Although theoretically the Eu$^{3+}$ ground state has zero magnetic moment, experimentally it exhibits an effective magnetic moment of 3.4 $ \mu_B$ due to Van Vleck paramagnetism \cite{van1968magnetic,takikawa2010van}. Therefore, the substitution of Eu at the La site can introduce an additional paramagnetic contribution and through modification of local structural environment can influence the magnetic properties of Cr$^{3+}$ sub lattice.  

In this work, we synthesised 5$\%$ Eu-doped LaCrO$_3$ (ELCO) and LaCrO$_3$ (LCO)  and carried out a detailed high-pressure and low temperature investigation using synchrotron X-ray diffraction (XRD), Raman spectroscopy, and magnetization measurement to elucidate the effect of pressure on the structural, vibrational and magnetic properties of the system. High-pressure XRD measurements reveal a structural phase transition in ELCO from orthorhombic to rhombohedral phase at 10.6 GPa. The Raman spectroscopic data also corroborate the structural transition and additionally exhibit an abrupt change in the frequencies of certain optical phonon modes around 4.5 GPa, accompanied by an anomaly in the  integrated intensity of the Raman modes. Similar anomalies in the phonon frequencies and integrated Raman intensities are observed in LCO at it's antiferromagnetic transition pressure. The low temperature magnetization measurements reveal the decrease of Neel temperature upon Eu doping. The pressure evolution of Neel temperature further indicates that the anomalies observed in the Raman data of ELCO at 4.5 GPa is associated with the onset of a long range antiferromagnetic ordering revealing the strong magneto-elastic coupling in the system under pressure.      
\section{Experimental details}
Polycrystalline samples of La$_{1-x}$Eu$_x$CrO$_3$ (x= 0 and 0.05) were prepared by a high-pressure high-temperature synthesis route using a piston cylinder anvil press \cite{xu2019high}. High-purity starting materials of La$_2$O$_3$, Eu$_2$O$_3$, and Cr$_2$O$_3$ were weighed in appropriate stoichiometric ratios and mixed well for 3 hours using an agate mortar. The mixed powders were enclosed in a platinum capsule and sintered at 3 GPa and 1623 K for 5 hours. Then the samples were quenched to room temperature, and the pressure was slowly released. The phase purity of the samples was confirmed by using lab XRD. The chemical compositions were confirmed by using energy-dispersive X-ray spectroscopy (EDX), as shown in Fig.S1 of the supplementary material. 

High-pressure XRD data for ELCO at room temperature were collected at the PETRA III P02.2 beamline, Germany, at the wavelength of 0.2907\AA with a spot size of (8$\times3$)~\SI{}{\micro\meter}$^2$. A symmetric diamond anvil cell (DAC) with 300~\SI{}{\micro\meter} culet was used for the high-pressure experiment. A rhenium gasket was preindented to 45~\SI{}{\micro\meter} thickness and a central hole of 100~\SI{}{\micro\meter} diameter was drilled using an electric discharge machine. Then the gasket was placed on the lower diamond, and the sample, along with a ruby sphere, was loaded into the central hole. The ruby sphere served as a pressure calibrant, and the pressure was determined by using the ruby fluorescence method \cite{mao1986calibration}. Neon gas was loaded into the gasket hole to serve as a pressure-transmitting medium (PTM). The XRD pattern of the standard sample CeO$_2$ was used to calibrate the sample to detector distance. The two-dimensional diffraction images were integrated to 2$\theta$ vs intensity profile using DIOPTAS software \cite{prescher2015dioptas}. The GSAS \cite{toby2001expgui}, VESTA \cite{momma2008vesta}, and  EoSfit7 \cite{gonzalez2016eosfit7} software were used for the analysis of XRD data.  

High-pressure Raman spectroscopic measurements were carried out using a piston-cylinder type DAC from Almax EasyLab Co. with 300~\SI{}{\micro\meter} culet. Ruby fluorescence technique \cite{mao1986calibration} was employed for pressure calibration, and 4:1 methanol- ethanol  mixture was used to maintain a hydrostatic environment inside the sample chamber. The Raman scattering measurements were performed using a confocal micro Raman spectrometer (Monovista from S\&I GmbH) in backscattering geometry with a 532 nm laser source. The laser was focused onto the sample using an infinitely corrected long working distance 20X objective, producing a spot size of around 4 \SI{}{\micro\meter}. An edge filter with a cutoff frequency of approximately 80 cm$^{-1}$ was used to cut the Rayleigh line, and a Bragg filter with cutoff frequency around 4 cm$^{-1}$ was used for Rayleigh line rejection, while collecting the low-frequency Raman modes. The magnetisation measurements were carried out by using a physical property measurement system (PPMS) with a magnetic field of 0.5 T.  
\section{Results and discussion}
\subsection{Ambient and high-pressure XRD}
\begin{figure}[ht!]
    \centering
    \includegraphics[width=1.0\linewidth]{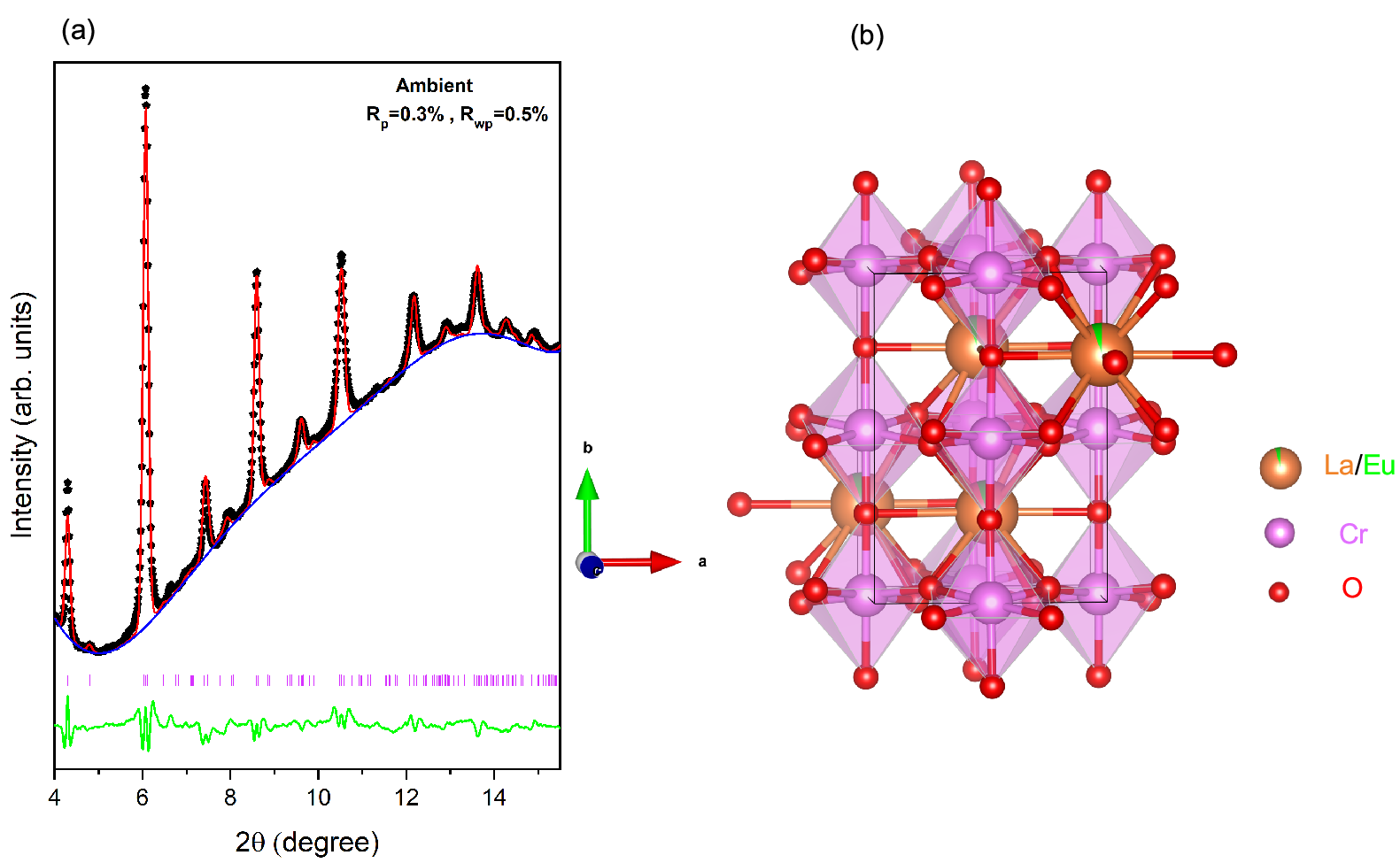}
    \caption{(a) The Rietveld refinement of the XRD pattern of ELCO at ambient pressure. The black dots are the experimental data points. The red line depicts the fit to the experimental data points and the blue line corresponds to the background. The green line represents the difference between the observed and fitted data. The pink coloured vertical lines show the Bragg reflection lines of the sample. (b) The representation of the unit cell of ELCO at ambient pressure. }
    \label{ambxrd}
\end{figure}

The ambient XRD patterns of ELCO is indexed using orthorhombic structure in the Pnma space group. The obtained lattice parameters from the best fit are a = 5.4597(1)\AA, b = 7.7442(2)\AA, c = 5.5341(9)\AA, and the unit cell volume $V_0$ = 233.99(1)\AA$^3$. The obtained unit cell volume of the LCO  is 234.489(4)\AA$^3$ and the lattice parameters are a= 5.4728(5)\AA, b= 7.7581(8)\AA, c= 5.5227(4)\AA. These values are in good agreement with the reported literature \cite{xu2019high,siddique2021intrinsic}. The smaller volume of the doped sample is due to the incorporation of Eu with a smaller ionic radius than La. The Rietveld refinement of the ambient XRD pattern of ELCO is shown in Fig.~\ref{ambxrd}(a), which gives a very good fit with $R_p = 0.3 \%$ and $R_{wp} = 0.5\%$. The refined atomic positions are tabulated in Table~\ref{atomic positions}.
\begin{table}[ht!]
    \centering
    \begin{tabular}{|c|c|c|c|c|}
    \hline
         Atoms&Wyckoff positions&x/a&y/b&z/c  \\
         \hline
         La/Eu&4c&0.0209(1)&0.25&0.0035(5)\\
         \hline
         Cr&4b&0.0&0.0&0.5\\
         \hline
         O(1)&4c&0.4980&0.25&-0.0430\\
         \hline
         O(2)&8d&0.2620&0.0320&-0.2820\\
         \hline
    \end{tabular}
    \caption{The refined atomic positions of ELCO at ambient pressure}
    \label{atomic positions}
\end{table}
 There are total 20 number of atoms in the unit cell of ELCO. The unit cell of ELCO consists of an array of corner shared tilted CrO$_6$ octahedra. The La/Eu cations occupy the interstitial sites of the  three-dimensional framework created by Cr$O_6$ octahedra and exhibit a twelvefold coordination with the O atom forming La/EuO$_{12}$ dodecahedra as shown in Fig.~\ref {ambxrd}(b).

\begin{figure}[ht!]
    \centering
    \includegraphics[width=0.8\linewidth]{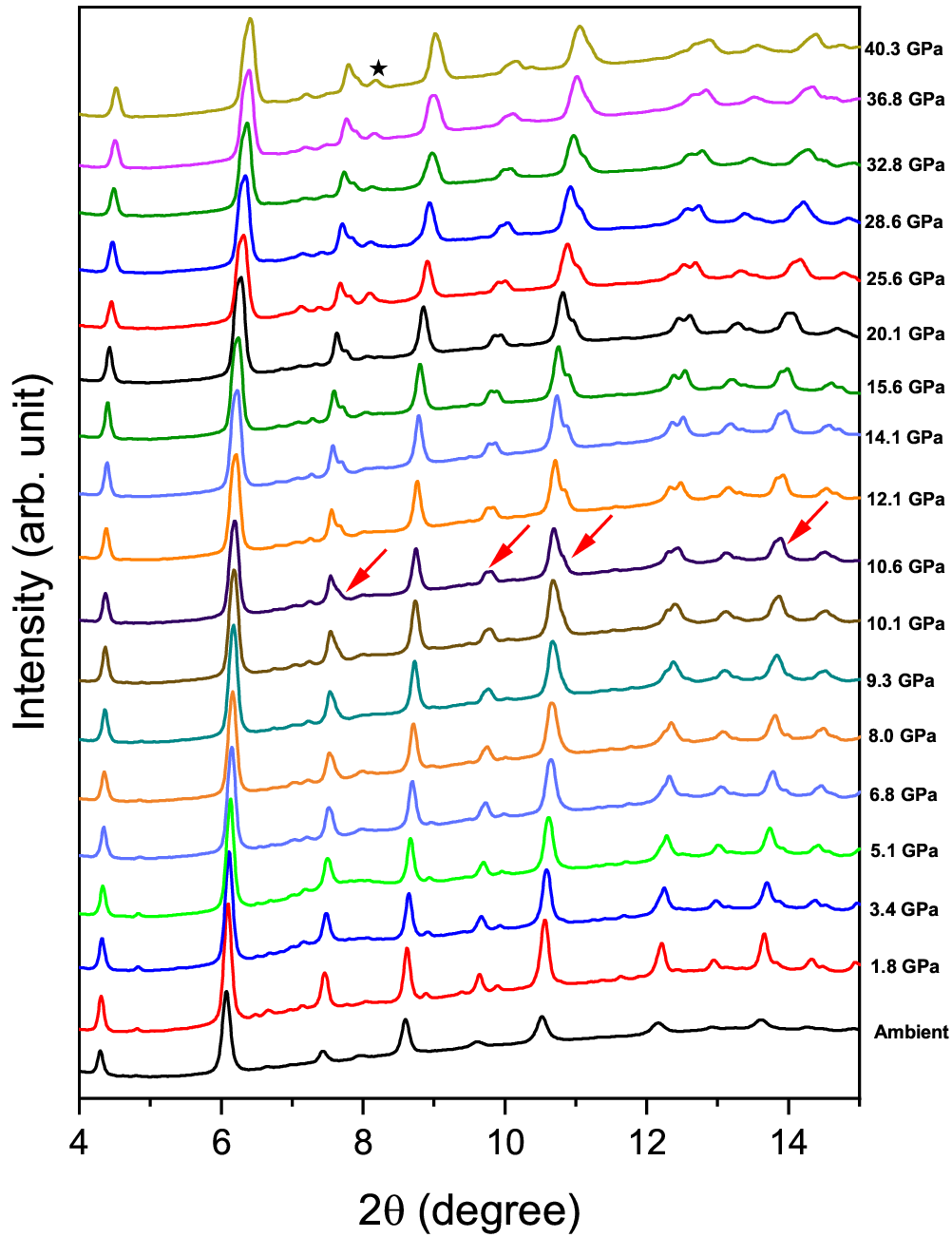}
    \caption{Pressure dependent XRD pattern of ELCO at selected pressures. The arrow marks point to the splitting of the Bragg peak, indicating structural phase transition. The Bragg peak marked with an asterisk corresponds to a small trace of Rhenium introduced into the sample chamber from the Rhenium gasket.   }
    \label{presure evolution xrd}
\end{figure}

\begin{figure}[ht!]
    \centering
    \includegraphics[width=1.0\linewidth]{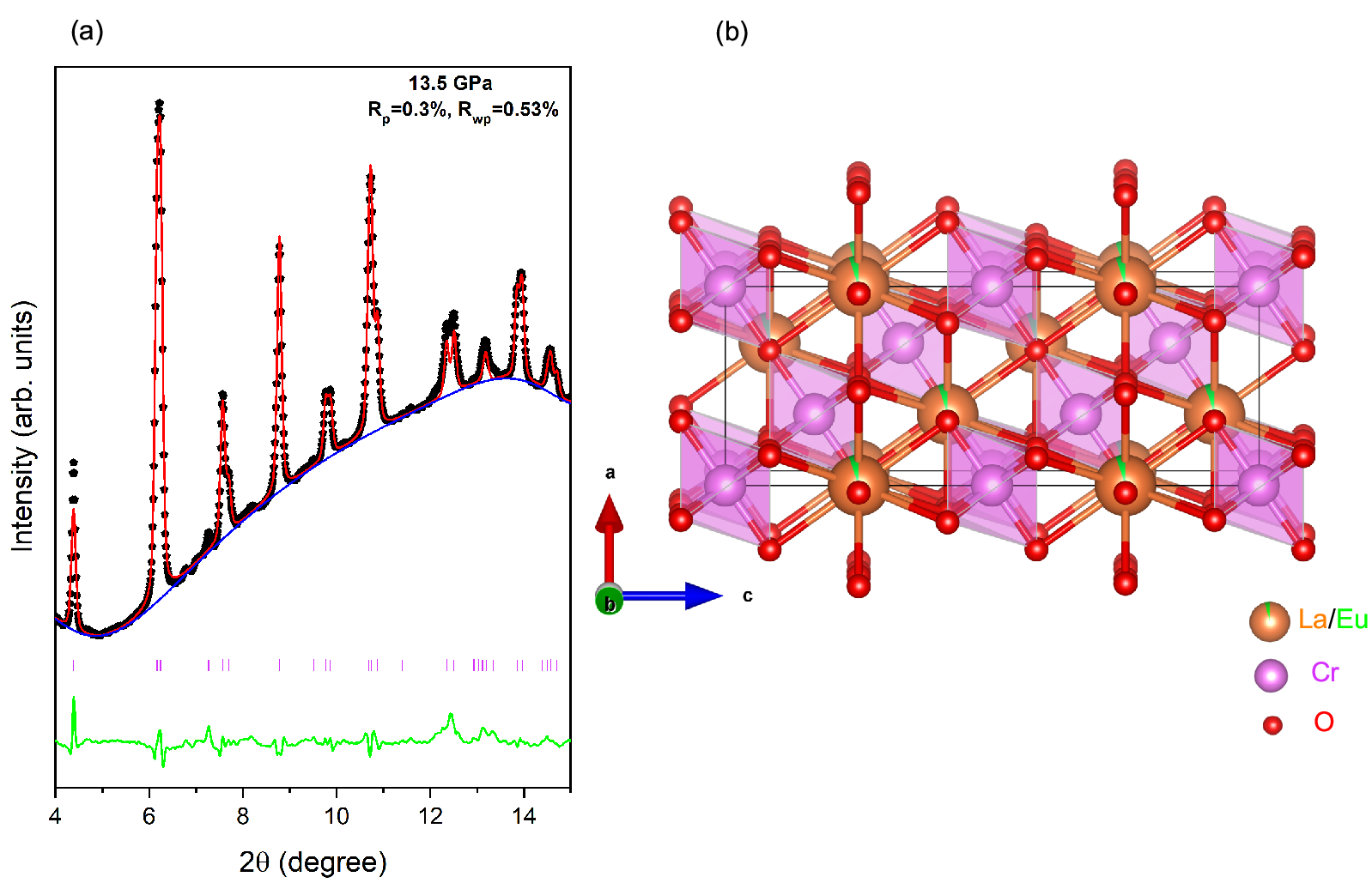}
    \caption{(a) The Rietveld refinement of the XRD pattern of ELCO at 13.5 GPa. The black dots are the experimental data points. The red line depicts the fit to the experimental data points and the blue line corresponds to the background. The green line represents the difference between the observed and fitted data. The pink coloured vertical lines show the Bragg reflection lines of the sample. (b) The representation of the unit cell of ELCO at 13.5 GPa.}
    \label{XRD_13.5GPa}
\end{figure}

\begin{figure}[ht!]
\centering
\includegraphics[width=1.1\linewidth]{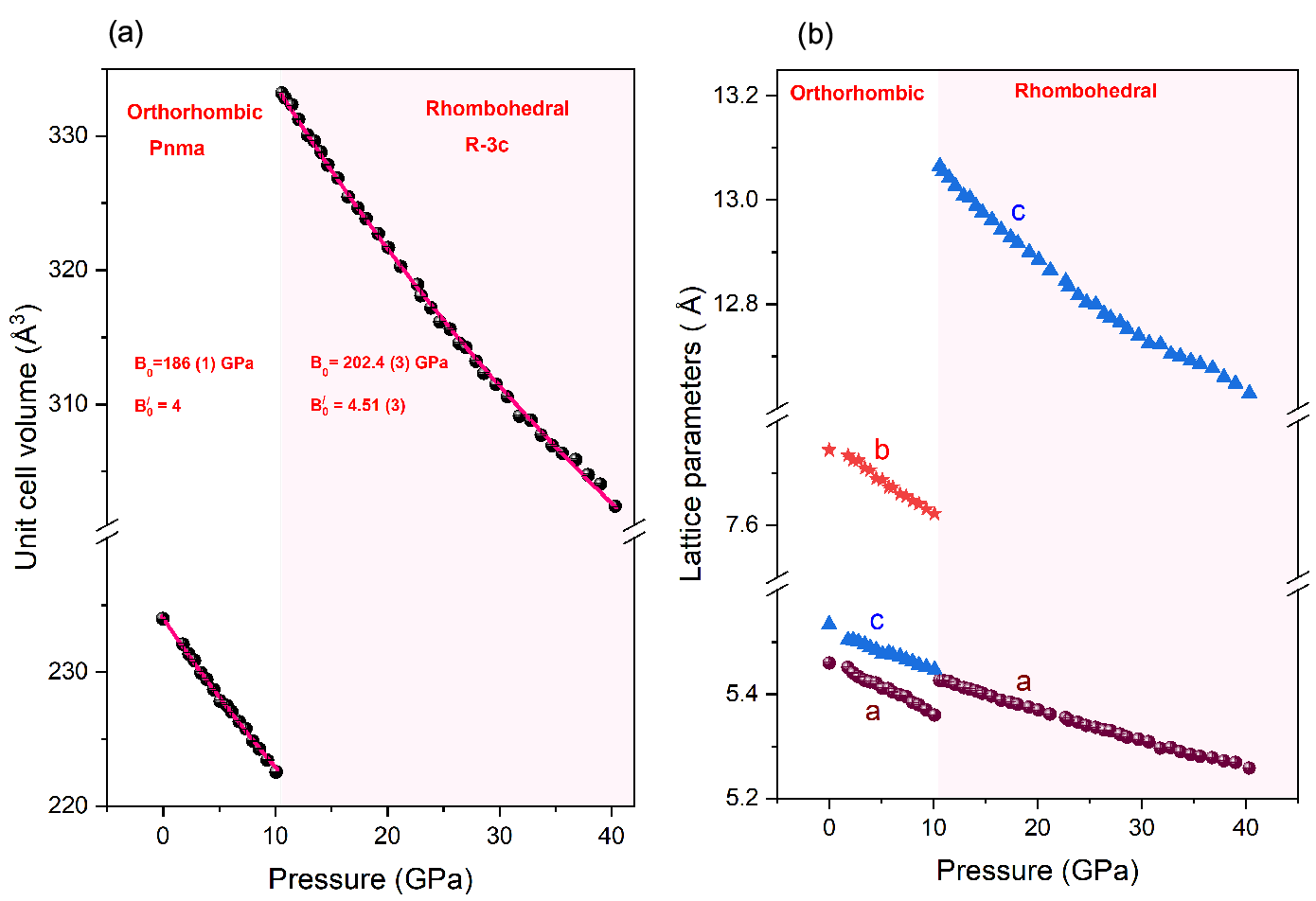}
\caption{(a) The pressure dependence of the unit cell volume of ELCO. The black spheres represent the experimental data and the red line is the 3rd order BM EOS fit to the experimental data. (b) The evolution  of lattice parameters as a function of pressure. The error bars are smaller than the symbol size.  }
\label{PV-latticparameter}	
\end{figure}

We have carried out the high-pressure XRD measurement of ELCO at different pressures up to 40.3 GPa. The XRD patterns collected at some selected pressure points is shown in Fig~\ref{presure evolution xrd}. With the increase in pressure the XRD patterns show no significant changes up to 10.6 GPa, indicating the stability of the ambient orthorhombic phase. At 10.6 GPa, certain Bragg peaks start splitting as indicated by arrows in Fig~\ref{presure evolution xrd}. With further compression, the splitting becomes well resolved and shows no further change up to the highest pressure studied. All the XRD patterns up to 10.6 GPa can be fitted well using the ambient Pnma space group. However, on and above 10.6 GPa, the Rietveld refinement does not give a good fit to the orthorhombic structure due to the splitting of several Bragg reflections. We would like to point out that the XRD experiments were carried out using Ne gas as PTM and hence the splitting of the Bragg peaks is not due to any non-hydrostatic stress. Therefore, we have re-indexed the XRD data at 10.6 GPa taking into account the additional Bragg reflections as marked by arrows in Fig.~\ref{presure evolution xrd}. It gives a rhombohedral structure with R-3c space group. The Rietveld refinement was performed by taking the initial atomic positions of rhombohedral phase of LCO, as reported by Oikawa et al. from neutron diffraction measurements \cite{oikawa2000structural}. The refined lattice parameters are a= 5.4266(4) \AA, c= 13.0654(2) \AA, and the unit cell volume is V= 333.20(5) \AA$^3$ at 10.6 GPa. The orthorhombic-rhombohedral phase transition has been reported for the parent compound LaCrO$_3$ but at a significantly lower pressure of $\sim$ 5 GPa \cite{zhou2011magnetic,bhadram2021pressure,hashimoto1998pressure}. Substitution of 5\% Eu at the La site stabilizes the orthorhombic tilt arrangement via ionic radius mismatch and therefore a higher pressure is required to change the crystal structure. The Rietveld refinement of the XRD pattern at 13.5 GPa is shown in the Fig.~\ref{XRD_13.5GPa}(a). The unit cell structure is modified with a layered zig-zag chain of CrO$_6$ octahedra in the ac plane. The unit cell volume as a function of pressure obtained after Rietveld refinement is fitted to 3$^{rd}$ order Birch-Murnaghan (BM) equation of state (EoS) \cite{birch1947finite,murnaghan1944compressibility} as shown in the Fig.~\ref*{PV-latticparameter}(a). 
 \begin{equation}
	P(V)= \frac{3B_0}{2}\left[(\frac{V_0}{V})^{7/3}-(\frac{V_0}{V})^{5/3}\right]\times\left\{1+\frac{3}{4}(B^\prime_0-4)\left[(\frac{V_0}{V})^{2/3}-1\right]\right \}
\end{equation}

\begin{figure}[ht!]
	\centering
	\includegraphics[width=1.0\linewidth]{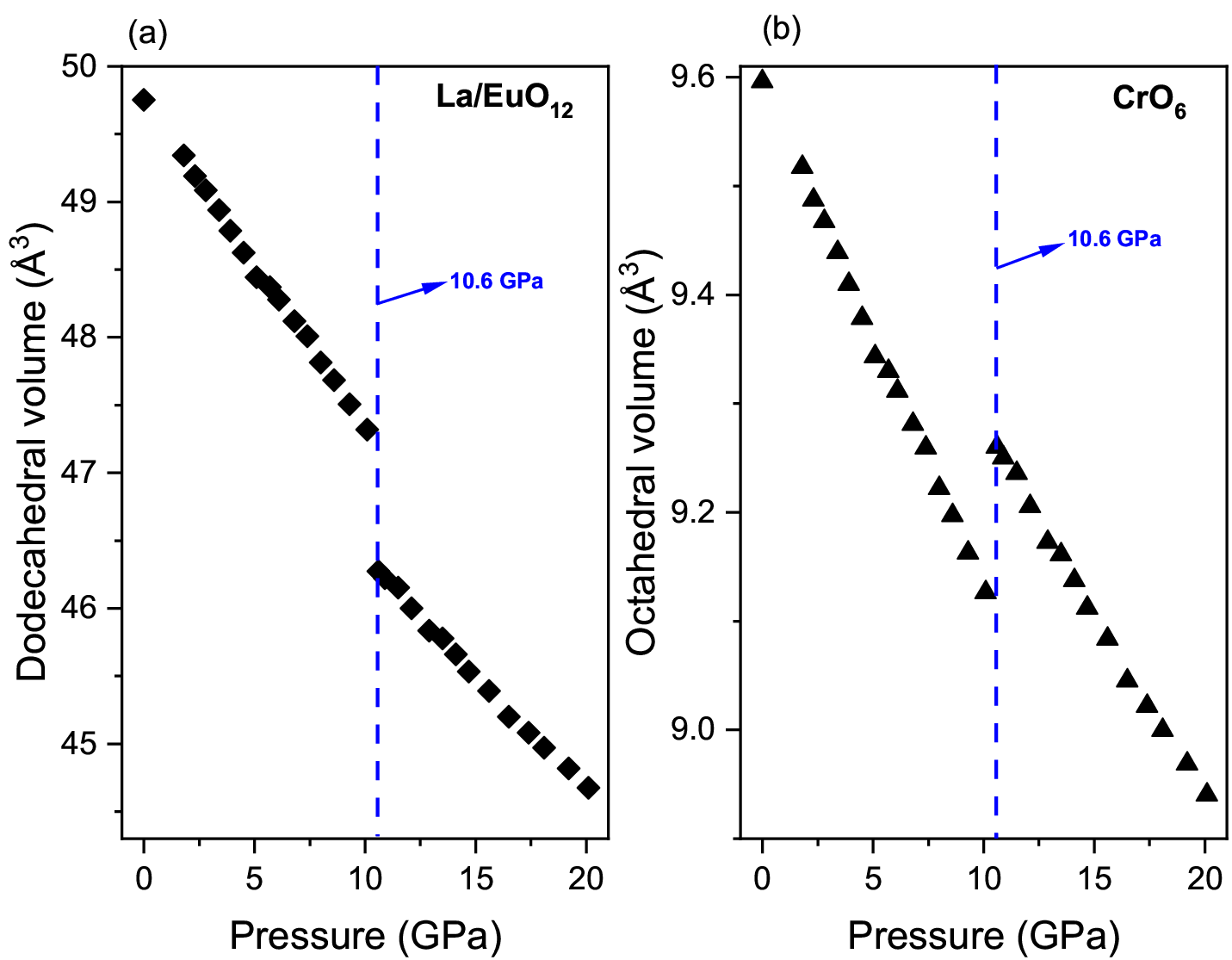}
	\caption{The pressure evolution of polyhedral volume of ELCO. The dashed vertical line corresponds to the orthorhombic to rhombohedral phase transition pressure. The error bars are smaller than the symbol size.   }
	\label{polyhedral volume}	
\end{figure}

Here, $B_0$ and $B^\prime_0$ represents the isothermal bulk modulus and  the first order pressure derivative of isothermal bulk modulus respectively, while $V_0$ is the unit cell volume at  ambient pressure. The obtained value of bulk modulus $B_0$ from the fitting are 186(1) and 202.4(3) for the orthorhombic and rhombohedral phases respectively, whereas the corresponding first order pressure derivatives $B^\prime_0$  are 4 (fixed) and 4.51(3) . Zhou et al. \cite{zhou2011magnetic} reported a bulk modulus of 188 for the orthorhombic phase of LCO. The close agreement between the $B_0$ values of ELCO and LCO in the orthorhombic phase indicates that a 5$\%$ Eu substitution at the La site has a negligible effect on the compressibility of the system under pressure. The transition from the low symmetry Pnma phase to high symmetry R-3c phase does not produce any noticeable discontinuity in the unit cell volume per formula unit (V/Z) as shown in the Fig.S2 of the supplementary material. The absence of such discontinuity in the V/Z data across the orthorhombic to rhombohedral transition pressure has also been reported for GdAlO$_3$ and LaGaO$_3$ \cite{mora2023experimental,kennedy2001pressure}. The pressure evolution of lattice parameters is  illustrated in Fig.~\ref{PV-latticparameter}(b). As discussed earlier, The unit cell of ELCO is composed of CrO$_6$ octahedra and La/EuO$_{12}$ dodecahedra. The evolution of polyhedral volumes at the transition pressure provides a further insight to the transition mechanism: At 10.6 GPa, corresponding to the onset of rhombohedral phase,  the La/EuO$_{12}$ dodecahedra undergo an abrupt decrease in volume, whereas the CrO$_6$ octahedra exhibits an volume expansion as shown in the Fig.~\ref{polyhedral volume}. A Similar behavior of the CrO$_6$ octahedra has been reported at the orthorhombic to rhombohedral phase transition of LCO under temperature. However, in that case, the LaO$_{12}$ dodecahedral volume does not show any anomaly at the transition temperature \cite{hashimoto2000analysis}. In contrast, for ELCO, the simultaneous compression of the La/EuO$_{12}$ dodecahedra, expansion of CrO$_6$ octahedra and a decrease in the octahedral tilt compensate for one another resulting a continuous behavior of V/Z across the pressure induced structural phase transition. As ELCO is a Jahn-Teller inactive system, the observed structural phase transition is primarily influenced by the pressure induced changes in the octahedral tilting, which leads to changes in the local site distortion of CrO$_6$ octahedra.  The local site distortion is described by O$_{21}$-Cr-O$_{22}$ bond angle, which is 90$^\circ$ for an ideal octahedron \cite{zhou2005universal,zhou2010intrinsic}. The pressure evolution of this bond angle is shown in Fig.~\ref{O-Cr-O bond angle}. Upon compression, initially this bond angle remains nearly constant up to 4.5 GPa and then increases monotonically up to 10.1 GPa, followed by an abrupt drop in it's value . This shows that the local site distortion decreases at the phase transition leading to pressure induced transition from low symmetry to high symmetry phase. The pressure independent behavior till 4.5 GPa in the orthorhombic phase is quite interesting and suggests a change in physical property of the system . To further investigate this anomaly and the effect of changes in octahedral rotation on the vibrational properties of the system, we have carried out a detailed Raman spectroscopic investigation under pressure.  
\subsection{High pressure Raman} 
\begin{figure}[ht!]
		\centering
		\includegraphics[width=0.9\linewidth]{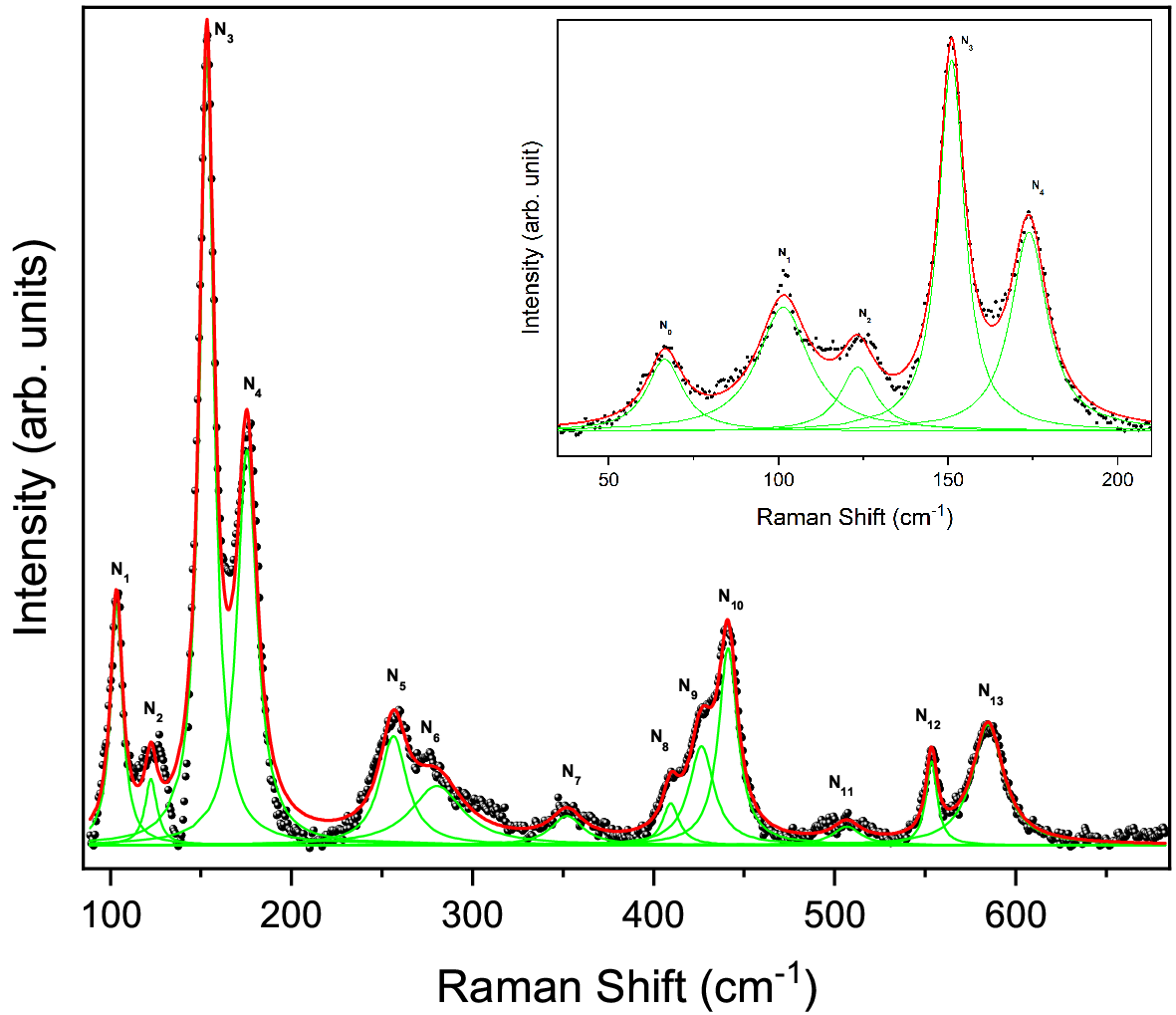}
		\caption{ The Raman spectra of ELCO at ambient pressure. The Lorentzian profile is used to fit the experimental spectrum after background correction. The Raman spectra in the main panel is collected, while using an Edge filter for Rayleigh line rejection and the inset shows the Raman spectra of ELCO in the low frequency region collected using an Bragg filter for Rayleigh line rejection   }
		\label{Amb_Raman}	
	\end{figure}

\begin{figure}[ht!]
		\centering
		\includegraphics[width=1.0\linewidth]{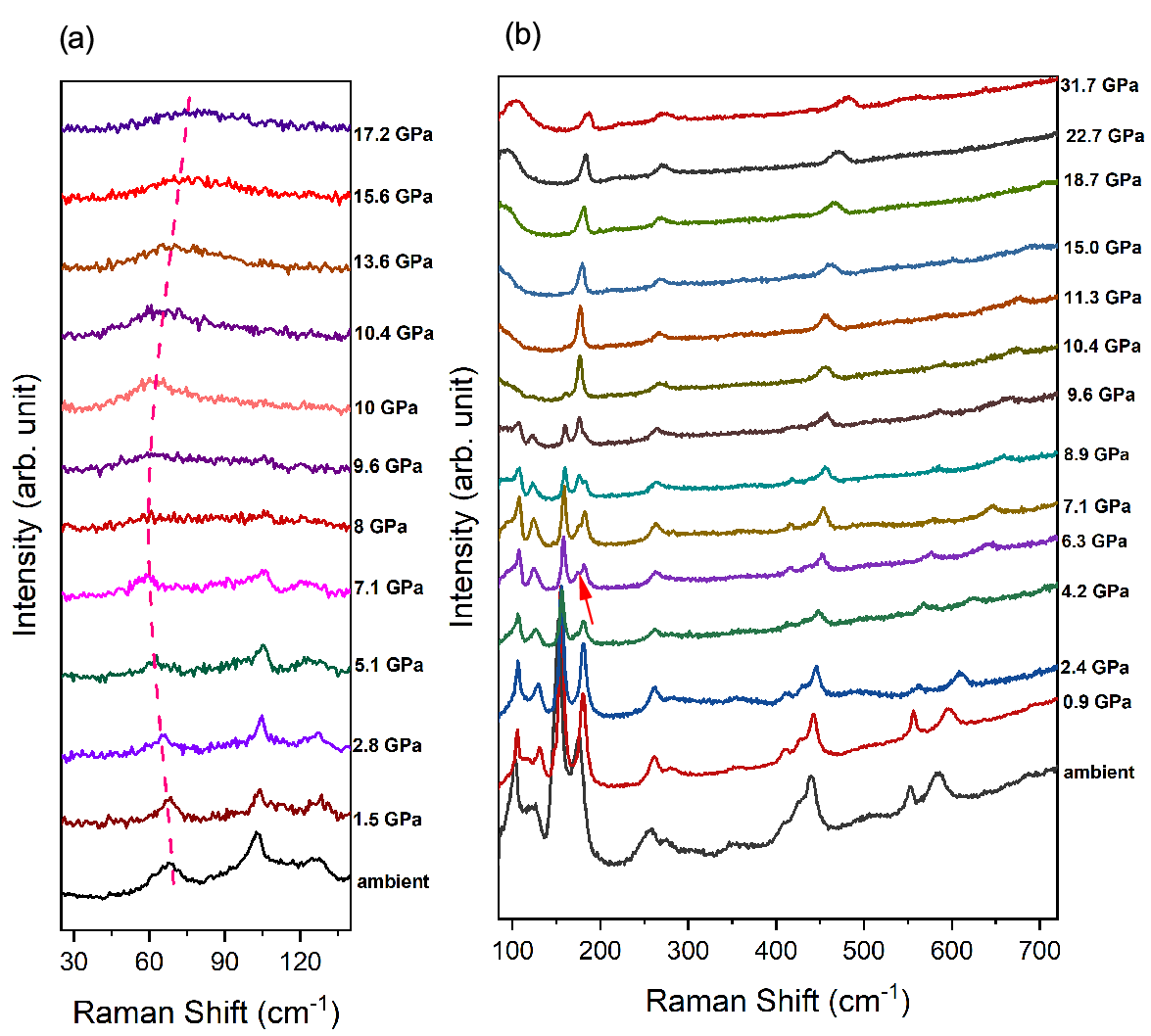}
		\caption{(a) The Pressure evolution of the Raman spectra collected using an Bragg filter . The pink colored dashed line is a guide to the eye for tracking the evolution of low frequency Raman mode under pressure. (b) The pressure dependent Raman spectra of ELCO,  collected by using an Edge filter. The arrow indicates the appearance of a low intensity Raman mode at 6.3 GPa   }
		\label{Raman_pressure evolution}	
	\end{figure}

  At ambient conditions, the group theory predicts 24 Raman active modes for orthorhombic Pnma structure \cite{iliev2006raman}. Experimentally, however, we observe a total of 14 Raman active modes in ELCO at ambient conditions. The ambient Raman spectra of ELCO is shown in Fig.~\ref{Amb_Raman}. To probe the low frequency Raman mode, the Raman spectra was also collected using a Bragg filter and the spectra is shown as an inset in the Fig.~\ref{Amb_Raman}. The observed Raman modes are assigned based on the mode assignments reported by Iliev et al. and Camara et al \cite{iliev2006raman,camara2017polarized}. The pressure evolution of the Raman spectra is presented in Fig.~\ref{Raman_pressure evolution}. At around 6.3 GPa, a new Raman mode with very low intensity starts appearing around 172 cm$^{-1}$ as marked by an arrow in Fig.~\ref*{Raman_pressure evolution}(b). with further increase in pressure, the intensity of this mode gradually increases as compared to other Raman modes. Above 10.4 GPa, the Raman spectral pattern undergoes a complete reconstruction marking the onset of Rhombohedral phase. Pressure evolution of the Raman mode frequencies are shown in the Fig.S3 of the supplementary material. At the structural phase transition, several Raman modes disappear, accompanied by changes in the remaining phonon modes as evident from the Fig.S3. However, we find certain anomalies in the frequencies of N$_0$, N$_1$, N$_2$, N$_3$ and N$_4$ Raman modes. Now we will discuss the behavior of these individual Raman modes under pressure in detail.
       
\begin{figure}[ht!]
	\centering
	\includegraphics[width=1.0\linewidth]{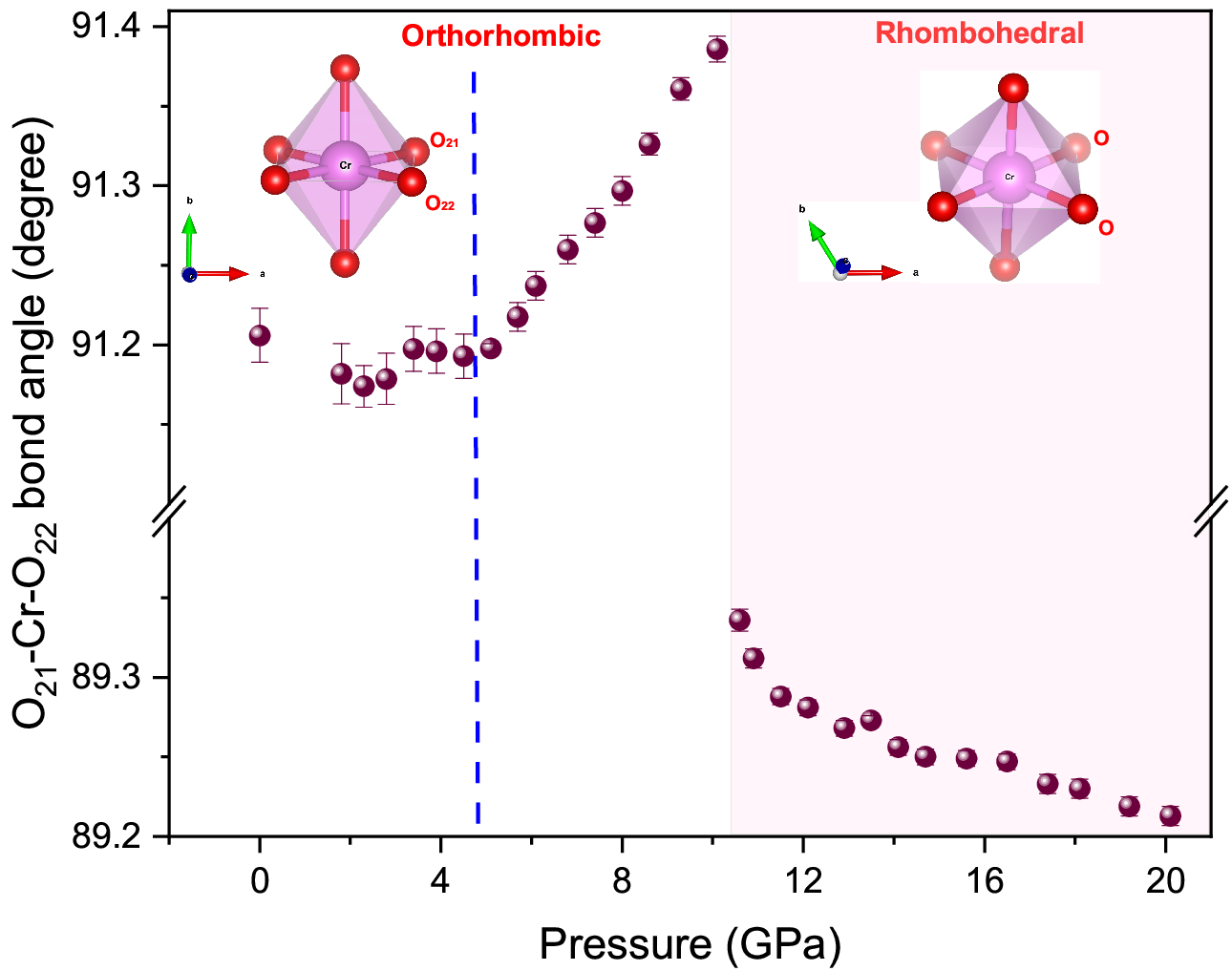}
	\caption{ The pressure dependence of O$_{21}$-Cr-O$_{22}$ angle . The vertical dashed line corresponds to 4.5 GPa  }
	\label{O-Cr-O bond angle}	
\end{figure}

\begin{figure}[ht!]
		\centering
		\includegraphics[width=1.1\linewidth]{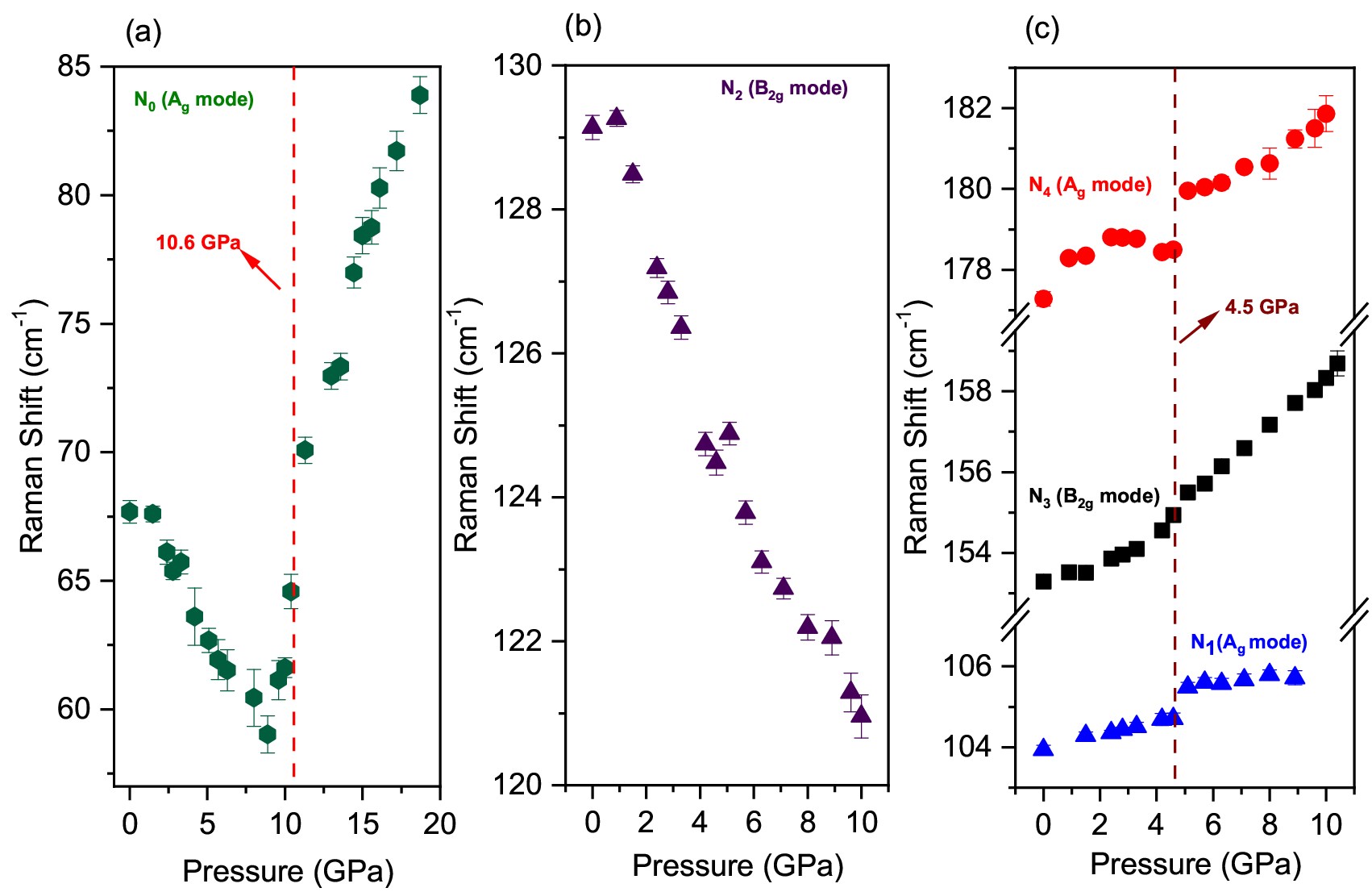}
		\caption{(a) The evolution of Raman Shift of low frequency N$_0$ mode under pressure. The vertical dashed line corresponds to structural phase transition pressure (b) The softening of the N$_2$ Raman mode under pressure (c) The pressure evolution of N$_1$, N$_3$ and N$_4$ Raman modes under pressure. The vertical dashed line corresponds to 4.5 GPa }
		\label{all mode}	
	\end{figure}

\begin{figure}[ht!]
		\centering
		\includegraphics[width=1.0\linewidth]{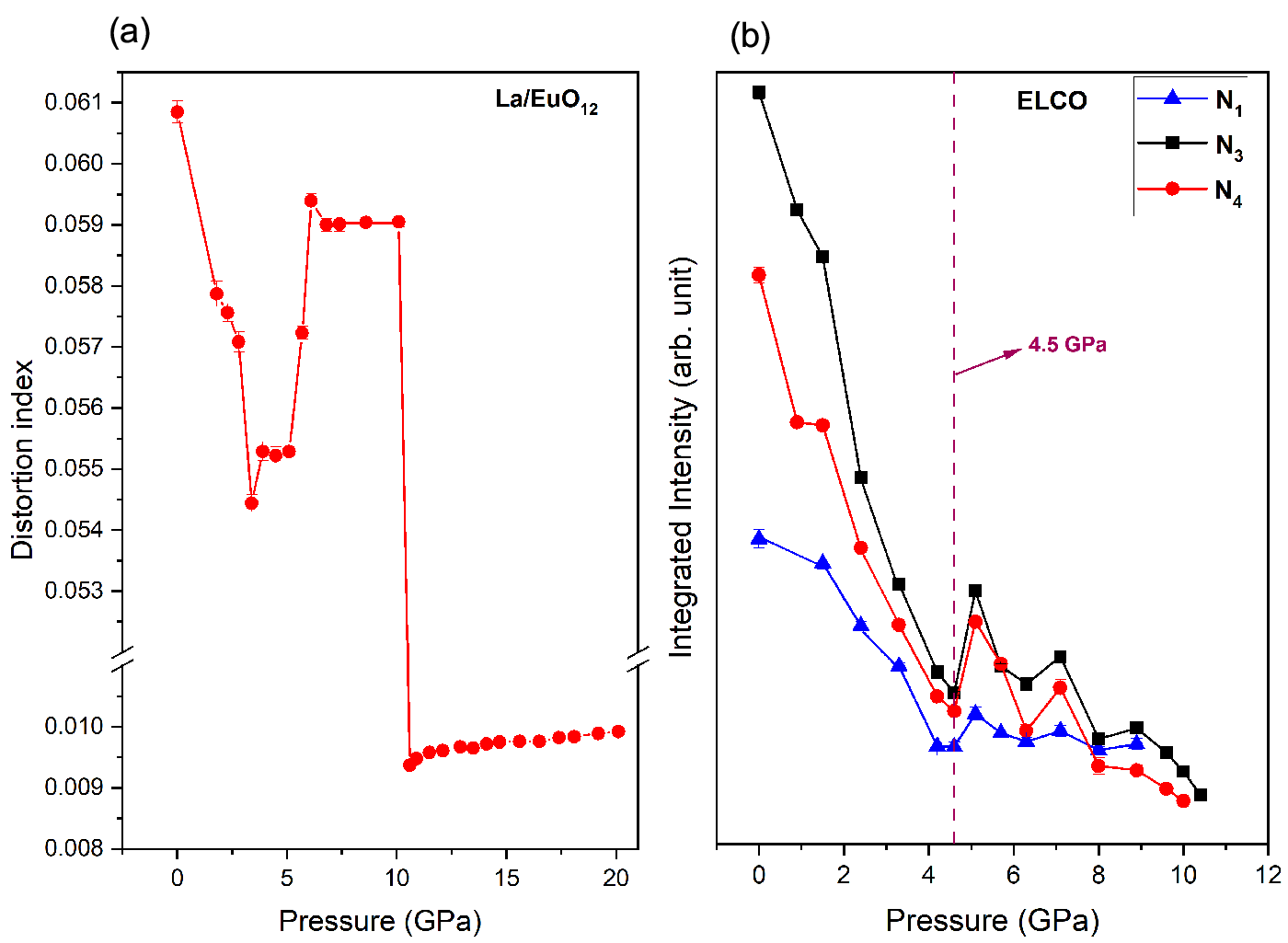}
		\caption{(a) Pressure evolution of the distortion index (DI) of (La/Eu)O$_{12}$ dodecahedra (b) Variation of integrated intensity of N$_1$, N$_3$ and N$_4$ Raman modes under pressure.   }
		\label{dist index LaEuO12}	
	\end{figure}

  The low frequency Raman mode at around 67 cm$^{-1}$ is identified as a A$_g$ Raman active mode, which involves the motion of rare earth ion and oxygen atom within (x,z) plane \cite{camara2017polarized}. This mode is not observed in most of the previous Raman studies, likely due to it's very low frequency. With the increase in pressure the mode exhibits an anomalous softening behavior up to $\sim$ 10 GPa followed by a hardening at higher pressures as shown in the Fig.~\ref{all mode}(a). In the high pressure R-3c phase this mode is assigned E$_g$ Raman active mode \cite{tompsett2004characterisation}. Similarily, the N$_2$ mode is identified as a B$_{2g}$ Raman active mode and arises from the displacement of rare earth and O atom, exhibits softening till 10 GPa and then disappears, which is shown in the Fig.~\ref{all mode}(b). This anomalous softening behavior of both the modes coincides with the Pnma to R-3c transition, as observed in our high pressure XRD data. The motion of the O in the (x,z) plane is associated with the Cr-O bond length, hence the change in CrO$_6$ octahedral environment under pressure can significantly affect the frequency of these Raman modes. The softening of Raman modes are generally associated with lattice instabilities and they are known to drive the structural phase transition, as initially suggested by Cochran and Anderson \cite{cowley2012soft,mukherjee2025soft,karmakar2022structural}. The pronounced softening of these two Raman modes under pressure provides evidence for the increasing instability in the lattice as the structural phase transition is approached. This behavior is also in accordance with the evolution of O-Cr-O bond angle under pressure (Fig.~\ref{O-Cr-O bond angle}), which exhibits an abrupt change at the structural phase transition. Since both of these Raman modes involve the motion of rare earth and oxygen atoms, the evolution of distortion index (DI) of (La/Eu)O$_{12}$ dodecahedra under pressure gives further useful insight into the transition mechanism and is shown in the Fig.~\ref{dist index LaEuO12}(a). The dodecahedral DI initially decreases with increase in pressure and exhibits a minimum around 4.5 GPa. With further increase in pressure, it increases and becomes constant as the system approaches the transition followed by a significant drop at the structural phase transition. Such anomalous behavior of DI is an indicative of increasing structural instablities within the lattice as the pressure is increased.   The reduction in the distortion of dodecahedra at the structural phase transition is expected as the system enters into a higher symmetric R-3c phase.  The anomalous softening of Raman modes and the changes in O-Cr-O bond angle and dodecahedral distortion indicate an increasing structural instability with pressure, which ultimately drives the system towards a structural phase transition. The hardening of the N$_0$ Raman mode and disappearance of N$_2$ Raman mode above $\sim$ 10 GPa is due to the change in crystal structure to rhombohedral phase. 
  
  Now let us discuss about the behavior of N$_1$, N$_3$, and N$_4$ Raman modes under pressure , which is represented in the Fig.~\ref{all mode}(c). The Raman mode at $\sim$104 cm$^{-1}$ is identified as a A$_g$ Raman active mode, that involves the in phase y-rotation of CrO$_6$ octahedra \cite{iliev2006raman}. With the increase in pressure the mode shows typical blue shift up to 4.5 GPa and then exhibits a discontinuous jump followed by further hardening at higher pressures before disappearing at around 10 GPa. Similarily, the N$_3$ mode (B$_{2g}$), which involves the motion of rare earth ions, exhibits a slope change at around 4.5 GPa and disappears in the rhombohedral phase. The N$_4$ mode (A$_g$) also shows an abrupt blueshift around 4.5 GPa before disappearing above 10 GPa. The integrated intensity of these Raman modes decrease rapidly up to 4.5 GPa, beyond which the rate of decrease becomes significantly slower as shown in the Fig.~\ref{dist index LaEuO12}(b). The anomalies observed at 4.5 GPa is quite interesting as the local site distortion also shows anomalies at the same pressure as shown in Fig.~\ref{O-Cr-O bond angle}. As the magnetic properties of the system is mainly controlled by CrO$_6$ octahedra, the changes in O-Cr-O bond angle and the anomalous behavior of N$_1$, N$_3$, and N$_4$ Raman modes at 4.5 GPa can possibly be related to a change in magnetic ground state. Tiwari et al. reported a similar anomaly in the tempearture dependent behavior of B$_{3g}$ Raman mode of LCO, where it exhibits an abrupt blue shift at the antiferromagnetic ordering temeprature \cite{tiwari2013dielectric}. For LCO, Zhou et al. reported an antiferromagnetic ordering above 2.2 GPa pressure \cite{zhou2011magnetic}. The high pressure Raman study on LCO by Bhadram et al. \cite{bhadram2021pressure} mainly focussed on the structural transition and the behavior of Raman modes at the antiferromagnetic ordering pressure was not discussed. This motivates us to perform a high pressure Raman study on LCO.  
  
\begin{figure}[ht!]
		\centering
		\includegraphics[width=1.0\linewidth]{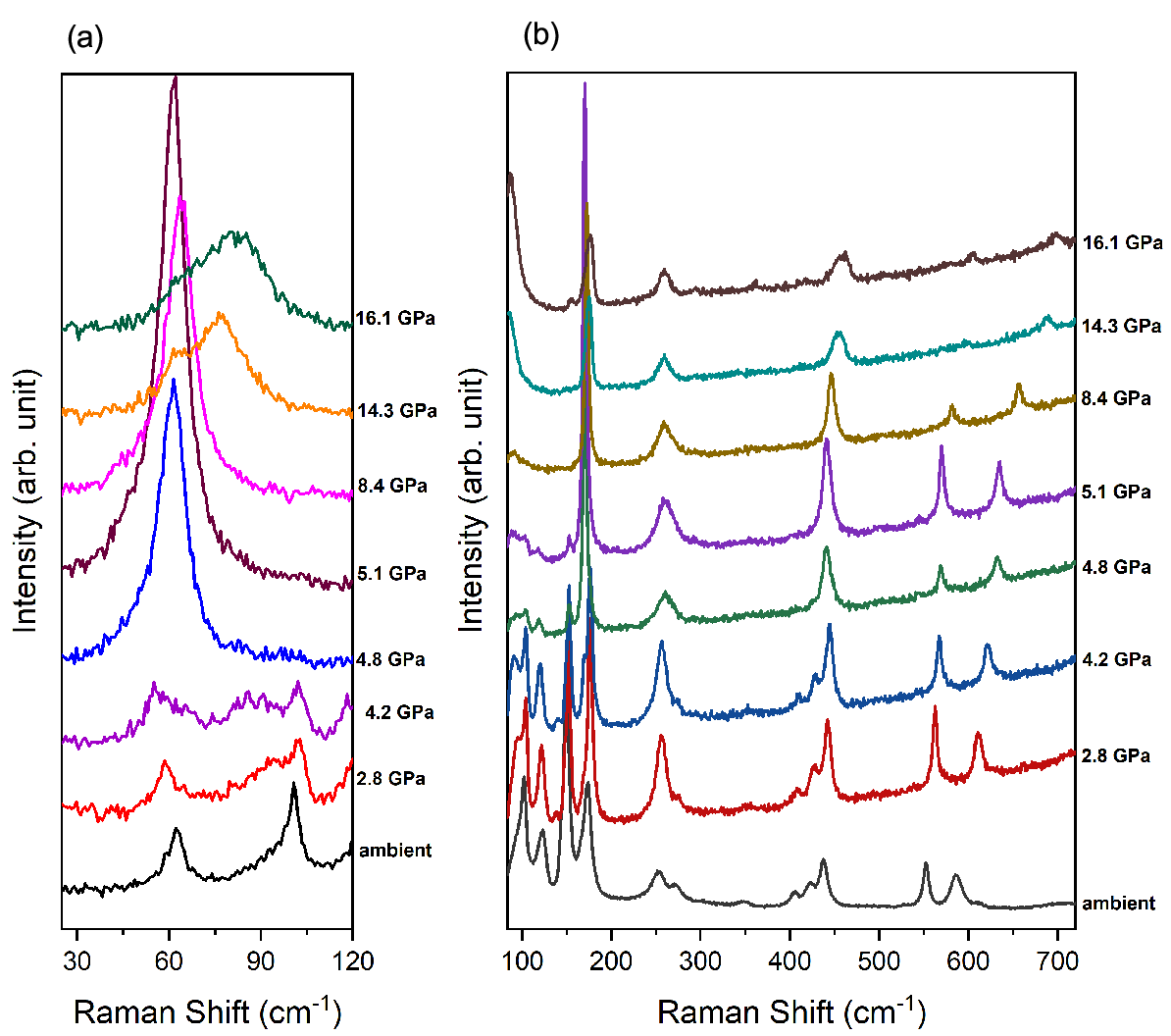}
		\caption{The pressure evolution of Raman spectra of LCO collected using (a) Bragg filter and (b) Edge filter  }
		\label{LCO_Raman_evolution}	
	\end{figure}

\begin{figure}[ht!]
		\centering
		\includegraphics[width=1.0\linewidth]{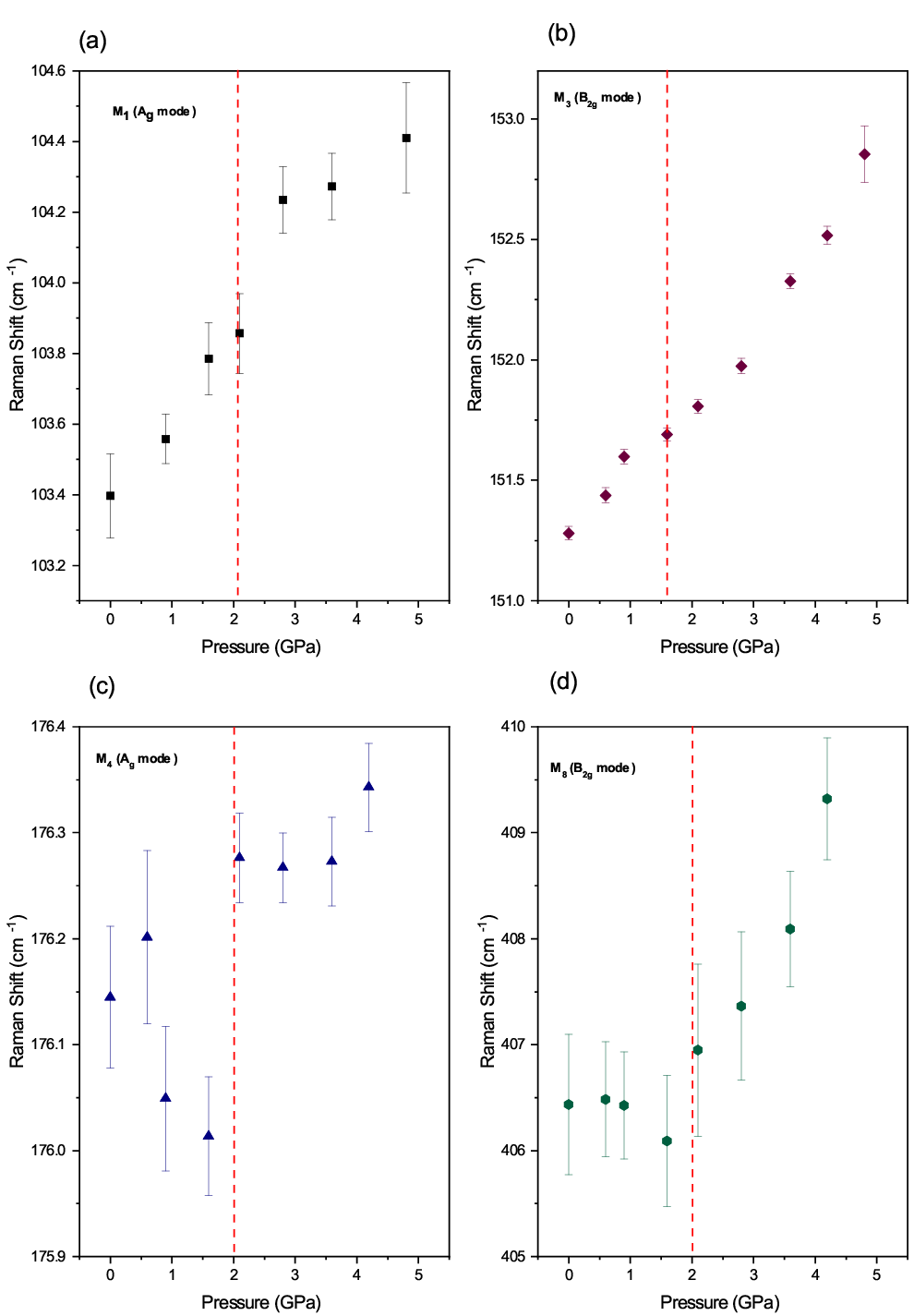}
		\caption{The variation of Raman peak position of LCO under pressure for (a) M$_1$ (b) M$_3$ (c) M$_4$ and (d) M$_8$ mode }
		\label{LCO_four_mode}	
	\end{figure}

  The ambient Raman spectra of LCO is shown in Fig.S4 of the supplementary material. With the increase in pressure, a new low intensity Raman mode appear around 169.6 cm$^{-1}$ at 4 GPa. This mode further intensifies with pressure and finally a reconstruction of Raman spectra occures around 5 GPa, as can be seen in the Fig.~\ref{LCO_Raman_evolution}. This signifies the structural phase transition from orthorhombic Pnma to rhombohedral R-3c structure. The transition pressure matches well with the reported literatures \cite{zhou2011magnetic,bhadram2021pressure}. The pressure dependent behavior of the Raman modes of LCO is shown in  Fig.S5 of the supplementary material. The pressure dependent behavior of the low frequency Raman mode M$_0$ has not been reported in the previous studies and is reported here for the first time. Similar to ELCO, we observe a pronounced softening in $M_0$ and $M_2$ Raman modes upon increasing the pressure. So the lattice instabilities associated with these two Raman modes are also involved in driving the structural phase transition in LCO. The pressure evolution of M$_1$, M$_3$, M$_4$ and M$_8$ Raman modes is shown in the Fig.~\ref{LCO_four_mode}. All these Raman modes exhibit an abrupt blue shift at the reported antiferromagnetic transition pressure along with an anomaly in the integrated Raman intensities (Fig.S6 of supplementary material), which strongly indicate towards the magneto-elastic coupling in the system. As we observe similar anomalies in the Raman mode frequencies of ELCO, we predict the possibility of a change in magnetic ground state in ELCO around 4.5 GPa.  
  
\begin{figure}[ht!]
		\centering
		\includegraphics[width=1.0\linewidth]{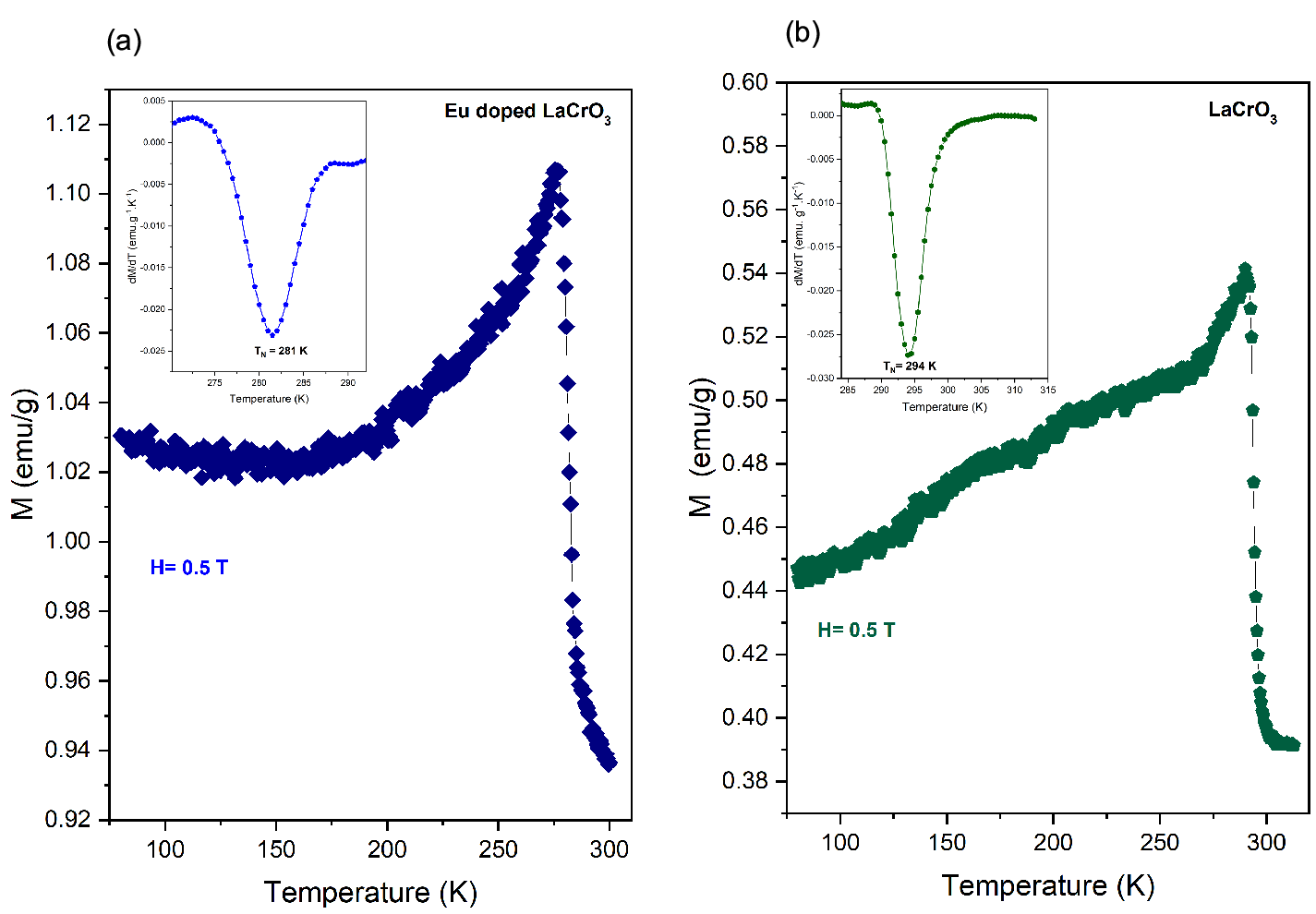}
		\caption{The magnetization as a function of temperature for (a) ELCO and (b) LCO under a magnetic field of 0.5 T. The inset shows the dM/dT vs temperature. }
		\label{MT}	
	\end{figure}

  For antiferromagnetic transition metal based oxides, Bloch gave an rule for the pressure dependent variation of magnetic transition temperature based on experimental results \cite{bloch1966103}. 
  \begin{equation}
  	  \frac{d (lnT_N)/dP}{k} \approx 3.3
  	  \label{Bloch_formula}
  \end{equation}    
  where, k is the compressibility and T$_N$ is the Neel temperature. The compressibility k can be found from the equation of state fit of the high pressure XRD data .  In order to know the exact value of T$_N$ , we carried out a low temperature magnetisation measurement for both LCO and ELCO sample. The magnetization vs temperature curves collected under a magnetic field of 0.5 T is shown in Fig.~\ref*{MT}.  A sharp enhancement in magnetization of both the samples at the magnetic transition temperature indicates the presence of weak ferromagnetism in the sample \cite{tiwari2015magnetostructural}. Zhou et al. reported the correlation between the structural and magnetic properties of RCrO$_3$ family  \cite{zhou2010intrinsic}. The local structural distortion causes the hybridization between the  intersite t$_{2g}$ and e$_g$ orbitals and enables a virtual charge transfer to the empty e$_g$ orbitals, which leads to the weak ferromagnetism in the system. The virtual charge transfer between the half filled t$_{2g}$ orbitals gives rise to the dominant antiferromagnetism in the system. The Eu doping modifies the octahedral site tilting, which in turn alters the orbital overlap between the t$_{2g}$ and e$_g$ resulting a shift in the Neel temperature. The observed value of Neel temperature for ELCO is 281 K and that of LCO is 294 K. 
  
  The Neel temperature is  calculated as a function of pressure by using the Eq.\ref{Bloch_formula} for both LCO and ELCO and is shown in the Fig.S7 of the supplementary material. The calculated Neel temperature increases with the increasing pressure and is predicted to exceed the room temperature near 2 GPa for LCO, and around 4.5 GPa for ELCO. Hence the observed anomalies in the Raman shift at this pressure can be attributed to the pressure induced transition from the paramagnetic to the antiferromagetic state via spin-phonon coupling.
  
  For magnetic materials the total Hamiltonian can be written as the sum of lattice and spin contributions, \[
  H=H_{Lattice}+H_{Spin},
  \]  As the studied systems exhibit an antiferromagnetic ordering, so by ignoring the higher order magnetic interaction terms ,the spin averaged Hamiltonian can be approximated as \begin{equation}
  	  H_{Spin}
  	\approx
  	-\sum_{i,j>i} J_{ij}
  	\left\langle \mathbf{S}_i\cdot\mathbf{S}_j\right\rangle
  \end{equation}
  where J$_{ij}$ represents the isotropic super exchange integral and $\left\langle \mathbf{S}_i\cdot\mathbf{S}_j\right\rangle$ is the spin correlation function of the nearest neighbour spins having spin angular momenta $\mathbf{S}_i$ and $\mathbf{S}_j$. When a long range magnetic ordering sets in, the phonons not only interact with the other phonons but also starts interacting with the spin of the electron resulting a strong spin-phonon coupling, which gives rise to a change in phonon self energy. By considering the nearest neighbour approximation, Granado et al. proposed a mechanism for the shift in phonon frequency due to magnetic ordering \cite{granado1999magnetic,issing2010composition}. According to this model, the shift in phonon frequency is given by \begin{equation}
  	\Delta\omega =
  	-\frac{1}{2\mu_\alpha\omega_\alpha}
  	\sum_{i,j>i}
  	\frac{\partial^2 J_{ij}}{\partial u_\alpha^2}
  	\left\langle \mathbf{S}_i \cdot \mathbf{S}_j \right\rangle
  	\label{Granado formula}
  \end{equation} 
  where $\mu_\alpha$ is the reduced mass of the phonon, $\omega_\alpha$ is the frequency of the $\alpha^{th}$ phonon mode and $u_\alpha$ is the displacement of O$^{2-}$ ion placed between the i$^{th}$ and j$^{th}$ Cr$^{3+}$ ions. So, any change in the strength of the exchange integral $J_{ij}$ and spin correlation $	\left\langle \mathbf{S}_i \cdot \mathbf{S}_j \right\rangle$ will be observed as a shift in the Raman frequency. Only those Raman active phonon modes whose atomic displacements significantly modulate the exchange interaction show strong spin-phonon coupling. At ambient condition, both the ELCO and LCO are in paramagnetic state. With the increase in pressure, the reduction of bond length and modification of octahedral geometry enhances the nearest neighbour and next nearest neighbour exchange interaction, thereby strengthening the spin correlation $	\left\langle \mathbf{S}_i \cdot \mathbf{S}_j \right\rangle$. Above a critical pressure of approximately 4.5 GPa for ELCO and 2 GPa for LCO the antiferromagnetic exchange interaction become sufficiently strong and leads to a long range ordering in the system. The onset of long range anti ferromagnetic ordering results in a non zero spin correlation function $	\left\langle \mathbf{S}_i \cdot \mathbf{S}_j \right\rangle$ and contributes to a positive $	\Delta\omega$ value as per the Eq.\ref{Granado formula}. This is manifested as an abrupt blue shift in the Raman mode frequencies as shown in Fig.~\ref{all mode}(c) and Fig.~\ref{LCO_four_mode}.
  
  The anomalous behavior of the integrated intensities of the Raman modes at the magnetic phase transition pressure can be understood by considering the spin dependent Raman scattering formalism developed by Suzuki and Kamimura \cite{suzuki1973theory}. The integrated intensity of the Raman mode in magnetic systems depends on the spin correlation function and is given by 
  \begin{equation}
  	I =\left|R+M\frac{\left\langle \mathbf{S}_i\cdot\mathbf{S}_j\right\rangle}{S^2}\right|^2
  	\label{Suzuki eq}
  \end{equation}
  where the first term R represents the spin independent contribution, while the second term represents the spin dependent contribution to Raman intensity. The parameters R and M are insensitive to the change in magnetic ordering. Although, the Suzuki-Kamimura formalism was originally developed to describe the temperature variation of Raman intensity in the magnetic systems, the same formalism can be used to describe the pressure induced anomalies in integrated Raman intensity. Zeng et al. used the above formalism to explain the changes in the Raman intensity across the magnetic transition of FePS$_3$ under pressure \cite{zeng2022raman}. Since the presently studied systems are in the paramagnetic state at the ambient pressure, so the contribution of spin independent part to the Raman intensity is expected to be dominant. As mentioned above, the increasing pressure strengthens the spin correlation function $\left\langle \mathbf{S}_i \cdot \mathbf{S}_j \right\rangle$. For a perfect antiferromagnetic state $\left\langle \mathbf{S}_i \cdot \mathbf{S}_j \right\rangle$ = $-\mathbf{S}^2$, which has a net effect of reducing the integrated Raman intensity of the system. At the magnetic transition pressure, the spin correlation function $\left\langle \mathbf{S}_i \cdot \mathbf{S}_j \right\rangle$ becomes non zero and negative due to the anti parallel alignment of the spins. This change in the spin correlation modifies the spin dependent contribution to the Raman intensity and is observed as an anomaly as shown in the Fig.~\ref{dist index LaEuO12}(b).  
 \section{Conclusion}    
            
  The 5$\%$ Eu doped LaCrO$_3$ undergoes a structural phase transition from the low symmetry Pnma phase to the high symmetry R-3c phase at around 10.6 GPa. A pronounced reduction in the distortion of both La/EuO$_{12}$ and CrO$_6$ octahedra is observed at the structural phase transition along with an anomaly around 4.5 GPa.  The high pressure Raman spectroscopic investigation corroborates the structural phase transition and reveals that the structural transition is driven by the softening of low frequency N$_0$ and N$_2$ Raman modes. An abrupt blue shift in the frequency of N$_1$, N$_3$ and N$_4$ Raman modes is observed around 4.5 GPa accompanied by a decrease in the integrated intensity of Raman modes. A similar Raman anomaly is  observed in the parent compound LCO at it's reported antiferromagnetic transition pressure of 2 GPa. By combining the low temperature magnetization measurements, Bloch formalism and the high pressure Raman results of LCO, the observed anomalies provide evidence for the onset of a long range antiferromagnetic ordering in ELCO above 4.5 GPa. The observed structural and vibrational anomalies around 4.5 GPa, well below the structural phase transition reveals the presence of strong magneto-elastic coupling in the system. 
  \section{Acknowledgments}
  Portions of this research were carried out at the light source PETRA-III of DESY, a member of the Helmholtz Association(HGF). Financial support by the Department of Science and Technology
  (Government of India) provided within the framework of the India@DESY collaboration is gratefully acknowledged. AKM acknowledges the IISER Kolkata, for the financial support and experimental facilities to carry out the PhD work.

\bibliography{name}
\bibliographystyle{unsrt}
\newpage
\section{figures}
	
	\end{document}